\pdfoutput=1
\documentclass{IEEEtran}

\usepackage{amsmath} 
\usepackage{amssymb}
\usepackage{bm} 
\usepackage{hyperref} 
\usepackage{tikz} 
\usepackage{tikzscale} 
\usepackage{gnuplot-lua-tikz}
\usepackage{listings} 

\newcommand{\Faust}{\textsc{Faust}}
\newcommand{\Faustsp}{\Faust{} }
\newcommand{\w}{\omega}
\newcommand{\wt}{{\tilde{\omega}}}
\newcommand{\wa}{\omega_a}
\newcommand{\Ht}{\tilde{H}}
\newcommand{\Gt}{\tilde{G}}
\newcommand{\Bt}{\tilde{B}}
\newcommand{\bt}{\tilde{b}}
\newcommand{\zt}{\tilde{z}}
\newcommand{\pt}{\tilde{p}}
\newcommand{\realpart}[1]{\mbox{re\ensuremath{\left\{#1\right\}}}}
\newcommand{\goestoas}[1]{\;\stackrel{\longrightarrow}{{\scriptscriptstyle #1}}\;}
\newcommand{\elabel}[1]{\label{e:#1}}
\newcommand{\eref}[1]{(\ref{e:#1})}
\newcommand{\pulse}{\sqcap}
\newcommand{\shah}{\mathbb{P}}
\providecommand{\zbox}[1]{\fbox{$\displaystyle #1$}}
\newcommand{\zi}{z^{-1}}
\newcommand{\ft}{\tilde{f}}
\newcommand{\fh}{\hat{f}}
\newcommand{\fmin}{f_{\mbox{min}}}
\newcommand{\fmax}{f_{\mbox{max}}}
\newcommand{\twobytwo}[4]{\left[\begin{array}{cc} #1 & #2 \\[2pt] #3 & #4 \end{array}\right]}
\newcommand{\twobyonenp}[2]{\begin{array}{c} #1 \\[2pt] #2 \end{array}}
\newcommand{\twobyone}[2]{\left[\twobyonenp{#1}{#2}\right]}
\newcommand{\rt}{\tilde{r}}
\newcommand{\ftmin}{\ft_{\mbox{min}}}
\newcommand{\ftmax}{\ft_{\mbox{max}}}
\newcommand{\Lscr}{{\cal L}}
\newcommand{\xdot}{x^{(1)}}
\newcommand{\xint}{x^{(-1)}}
\newcommand{\eqsp}{\;=\;}
\newcommand{\corrTo}{\;\longleftrightarrow\;}
\newcommand{\conv}{\;\ast\;}
\newcommand{\ints}{\mathbb{Z}}
\newcommand{\reals}{\mathbb{R}}
\newcommand{\beqn}{\begin{equation}}
\newcommand{\eeqn}{\end{equation}}
\newcommand{\beqa}{\begin{eqnarray}}
\newcommand{\eeqa}{\end{eqnarray}}
\newcommand{\beas}{\begin{eqnarray*}}
\newcommand{\eeas}{\end{eqnarray*}}
\newcommand{\ie}{\textit{i.e.}}
\newcommand{\eg}{\textit{e.g.}}
\newcommand{\Fref}[1]{Figure~\ref{fig:#1}}
\newcommand{\fref}[1]{Fig.~\ref{fig:#1}}
\newcommand{\ftu}[1]{\footnote{\url{#1}}}
\newcommand{\funcalign}[4]{\left\{\begin{array}{ll}
	#1, & #2 \\[5pt]
	#3, & #4 \\
	\end{array}
	\right.}

\providecommand{\isdef}{\mathrel{\stackrel{\mathrm{\Delta}}{=}}}
\providecommand{\isdefs}{\;\isdef\;}

\begin{document}

\title{Closed Form Fractional Integration and Differentiation via
  Real Exponentially Spaced Pole-Zero Pairs}

\author{Julius Orion Smith and Harrison Freeman Smith
\thanks{J. O. Smith is with CCRMA, Stanford University, CA, USA. 
e-mail: jos at ccrma.stanford.edu, website: https://ccrma.stanford.edu/\~{}jos/.}
\thanks{H. F. Smith has been studying mathematics at the University of
  Denver, CO, and will be an entering freshman at CU Boulder in the fall. e-mail:
  hfreemansmith at gmail.com.}
\thanks{Manuscript received June 5, 2016}
}


\maketitle

\begin{abstract}
We derive closed-form expressions for the poles and zeros of
approximate fractional integrator/differentiator filters, which
correspond to spectral roll-off filters having any desired log-log
slope to a controllable degree of accuracy over any bandwidth.  The
filters can be described as a uniform exponential distribution of
poles along the negative-real axis of the $s$ plane, with zeros
interleaving them.  Arbitrary spectral slopes are obtained by sliding
the array of zeros relative to the array of poles, where each array
maintains periodic spacing on a log scale.  The nature of the slope
approximation is close to Chebyshev optimal in the interior of the
pole-zero array, approaching conjectured Chebyshev optimality over all
frequencies in the limit as the order approaches infinity.  Practical
designs can arbitrarily approach the equal-ripple approximation by
enlarging the pole-zero array band beyond the desired frequency band.
The spectral roll-off slope can be robustly modulated in real time by
varying only the zeros controlled by one slope parameter.  Software
implementations are provided in matlab and \Faust.
\end{abstract}

\section{Introduction}

The notion of a \emph{fractional derivative} or \emph{integral} is
naturally defined in terms of the integration and differentiation
theorems for Laplace/Fourier transforms.  Let $X(s)$ denote the
\emph{bilateral Laplace transform} of $x(t)$:
\[
X(s) \isdefs \Lscr_s\{x\} \isdefs \int_{-\infty}^{\infty} x(t) e^{-st}dt,
\]
where $s=\sigma+j\omega$ is a complex variable, $t$ typically denotes
time in seconds, `$\isdef$' means ``equals by definition,'' and we
assume $x(t)$ and all of its integrals and derivatives are absolutely
integrable and approach zero as $t\to\pm\infty$.  Then the
\emph{differentiation theorem} for bilateral Laplace transforms states
that
\[
\Lscr_s\left\{\xdot\right\} 
\isdefs \Lscr_s\left\{\frac{d\,x(t)}{dt}\right\}
\eqsp s\,X(s)
\]
where $x^{(n)}(t)$ denotes the $n$th derivative of $x(t)$ with respect
to $t$.\footnote{In engineering, it is more common to use the
  \emph{unilateral Laplace transform}, in which the integral traverses
  $[0,\infty)$, and the differentiation theorem becomes
    $\Lscr_s\{\xdot\} = s\,X(s)-x(0)$.}  The proof is quickly derived
  using integration by parts.\ftu{https://ccrma.stanford.edu/~jos/filters/Differentiation.html}

The \emph{integration theorem} for Laplace transforms follows as a corollary:
\[
\Lscr_s\left\{\xint\right\} \isdefs \Lscr_s\left\{\int_{-\infty}^{t} x(\tau)\,d\tau\right\} \eqsp \frac{1}{s} \, X(s)
\]
The Laplace transform specializes to the Fourier transform along the
$s=j\omega$ axis in the complex plane, where $\omega = 2\pi f$ is
\emph{radian frequency} (radians per second), while $f$ denotes
frequency in Hz (cycles per second).

\subsection{Fractional Derivatives and Integrals}

Since the $N$th successive time derivative/integral of $x(t)$
Laplace-transforms to $s^{\pm N} X(s)$, it follows that the
\emph{fractional derivative} of order $\alpha\in(0,1)$, denoted by
$x^{(\alpha)}$, should correspond to the Laplace transform $s^\alpha
X(s)$, while the \emph{fractional integral} of order $\alpha\in(0,1)$
corresponds to $s^{-\alpha} X(s)$.  We thus let $x^{(\alpha)}$,
$\alpha\in\reals$, denote both fractional integrals ($\alpha<0$) and
derivatives ($\alpha>0$), possibly including both an integer and
fractional part.  However, since integer $\alpha=\pm N$ corresponds to
a repeated ordinary derivative or integral, we will henceforth
consider only $\alpha\in(-1,1)$ and implement any integer part in the
usual way.  Expressing $j$ as $e^{j\pi/2}$ so that $j^\alpha$ can be
defined as $e^{j\alpha\pi/2}$, we obtain the corresponding Laplace and
Fourier transforms for fractional integration or differentiation as
\beqn
\zbox{x^{(\alpha)}(t) \corrTo s^\alpha\, X(s) \;\goestoas{s=j\omega}\; e^{j\alpha\frac{\pi}{2}}\, \omega^\alpha\, X(j\omega).}
\elabel{fracfilt}
\eeqn
Thus, a fractional integral or derivative of order $\alpha$
corresponds to a phase shift by $\alpha \pi/2$ and a spectral ``tilt''
by $\omega^\alpha$.  For $\alpha=1$, we obtain the frequency-response
$j\omega$ of a differentiator, and for $\alpha=-1$, the integrator
frequency-response $-j/\omega$ is obtained, as required.

For integer $\alpha=N\in\ints$, $N\ge 0$, we have, from the
\emph{convolution theorem} for the \emph{unilateral} Laplace transform
applied to \emph{causal} functions $x(t)$,\footnote{A function of time
  $f(t)$ is said to be \emph{causal} if $f(t)=0$ for all $t<0$.}
\beqa s^{-N} X(s) &\longleftrightarrow& \frac{u(t)\,t^N}{(N-1)!} \conv
x(t)\nonumber\\[5pt] &=& \frac{1}{(N-1)!} \int_{0}^t x(\tau)\,
(t-\tau)^{N-1} d\tau.\elabel{iterint} \eeqa
where `$\ast$' denotes convolution, and $u(t)$ denotes the Heaviside
\emph{unit step} function:
\[
u(t) \isdefs \funcalign{1}{t\ge 0}{0}{t<0}
\]
This form was evidently developed originally as \emph{Cauchy's
  repeated integral
  formula.}\ftu{https://en.wikipedia.org/wiki/Cauchy_formula_for_repeated_integration}
\beas
x^{(-N)}(t) &\isdef& \int_a^{t} \int_a^{\tau_1} \cdots \int_a^{\tau_{N-1}} x(\tau_N)
		    \,d\tau_N\,d\tau_{N-1}\,\cdots d\tau_1\\[5pt]
&=& \frac{1}{(N-1)!}\int_a^{t} (t-\tau)^{N-1} x(\tau)\, d\tau
\eeas
where $a$ is any finite real number such that $x(t)=0,\,\forall t<a$.

The generalization of $s^{-N}$ to $s^\alpha$ for $\alpha\in(-1,1)$ is
quite natural. There is also no problem extending $(t-\tau)^{N-1}$ to
$(t-\tau)^{-\alpha-1}$ in \eref{iterint}, and the lower limit of
integration $a$ can be extended (for all practical purposes) 
as far as needed to the left to encompass the support of $x(t)$.  The
last piece is generalizing $(N-1)!$ to $(-\alpha-1)!$, which is
provided by the \emph{gamma
  function}\ftu{https://en.wikipedia.org/wiki/Gamma_function}
\[
\Gamma(t) \isdefs \int_0^\infty \tau^{t-1} e^{-\tau} d\tau
\]
which, for positive integers $N$, becomes $(N-1)$
factorial, \ie, 
\[
\Gamma(N)=(N-1)!
\]
for $N=1,2,3,\ldots\,$. We thus obtain the expression for fractional
integrodifferentiation in the time domain as the convolution integral
\[
x^{(\alpha)}(t) \isdefs \frac{u(t)\,t^{-\alpha}}{\Gamma(-\alpha)} \conv x(t)
\eqsp \frac{1}{\Gamma(-\alpha)} \int_0^t \frac{x(\tau)}{(t-\tau)^{\alpha+1}}d\tau.
\]
This is known as the \emph{Riemann-Liouville
  differintegral},\ftu{https://en.wikipedia.org/wiki/Differintegral}
more commonly stated closer to the following form:
\[
x^{(\alpha)}(t) = \frac{1}{\Gamma(-\alpha)} \int_a^t \frac{x(\tau)}{(t-\tau)^{\alpha+1}}d\tau.
\]
where $a$ is an arbitrary fixed base point, and $\alpha$ is any
complex number with $\realpart{\alpha}<0$.\footnote{Our definition of
  $\alpha$ is the negative of its usual definition in fractional
  calculus.  We keep it so that $\alpha$ can refer to the \emph{slope}
  of the Bode magnitude plot, instead of the negative slope.  Note
  also that we allow $\alpha>0$, but only consider $\alpha\in\reals$.}

The topic of fractional differentiation and integration falls within
the well studied subject of \emph{fractional
  calculus}\ftu{http://nrich.maths.org/1369}
\cite{OldhamAndSpanier74}.  We will adopt the term ``differintegral''
from that literature.

\subsection{Filter Interpretation}

As derived in obtaining \eref{fracfilt} above, every fractional
differintegral corresponds to a linear time-invariant \emph{filter}
having frequency-response
\beqn 
\zbox{H_\alpha(j\omega) \isdefs e^{j\alpha\frac{\pi}{2}} \, \omega^\alpha.}
\elabel{Halpha}
\eeqn
Since this frequency response is not a rational polynomial in
$j\omega$ for non-integer $\alpha$, there is no exact realization as
a finite-order filter \cite{JOSFP}.  We must therefore settle for a
finite-order approximation obtained using a truncated series expansion or
\emph{filter design}
technique \cite{ParksAndBurrus,JOST}.  Many filter-design methods are
available in the Matlab Filter Design
Toolbox,\ftu{https://www.mathworks.com/} and
several basic design methods, such as \texttt{invfreqz}, are also
available in the free, open-source, GNU Octave
distribution.\ftu{http://www.gnu.org/software/octave/}
As far as we know, all filter-based approximations to date have been
carried out along these lines.

\subsection{Exponentially Distributed Real Pole-Zero Pairs}

In contrast to exact integral forms or general-purpose filter designs
for fractional differintegrals, we will develop them in \emph{closed form}
as \emph{exponentially distributed pole-zero pairs}. (On a log scale, the
poles and zeros are \emph{uniformly} spaced.)  It appears that such
filters approach the Chebyshev optimal approximation (in terms of
log-log \emph{slope error}) for any $\alpha$ as the pole-zero density
and span along the negative real axis increase.

Any needed integer part $\pm M$ for $\alpha$ can be trivially provided
using $M$ zeros or poles at/near the origin $s=0$ of the complex
plane.

The proposed filter structure is furthermore robust for \emph{time-varying}
$\alpha$, because the poles are fixed, and only the zeros need to slide
left or right along the negative real axis of the $s$ plane as
$\alpha$ is changed. On a log scale, the spacing of the zeros does not
change as they are slid left and right.

The fractional order $\alpha$ sets the spacing of the zeros array
relative to the poles array along the negative-real axis.  For
$\alpha=1/2$, the zeros lie on the midpoints between the poles.  For
$\alpha=0$, the array of zeros slides to right so as to cancel all of
the poles, leaving the trivial filter $H_0(s)=1$, as desired.  At
$\alpha=-1$, all poles are canceled except the first to the left along
the negative real axis, leaving a normal integrator $H_{-1}(s)=1/s$,
as desired.  At $\alpha=+1$, all poles are canceled except the last,
and one zero is exposed near $s=0$, yielding a normal ``leaky
differentiator with high-frequency leveling''
$H_{1}(s)=p_N(s+\epsilon)/(s+p_N)\approx j\omega$ for frequencies
interior to the interval $\omega\in[\epsilon,|p_N|]$.  In the limit as
the number of poles $N$ goes to infinity, and as $\epsilon\to 0$, we
obtain the ideal differentiator $H_1(s)=s$, as desired.

\subsection{Importance in Audio Signal Processing}

In audio signal processing, we often need a spectral shaping filter
having a particular \emph{roll-off}, usually specified in decibels
(dB) per octave over the audio band. For example, it can be desirable
to arbitrarily set the \emph{slope} of the log-magnitude response
versus log frequency between the two transition frequencies of a
\emph{shelf filter} \cite{JOSFP}.

A more classical example is the synthesis of \emph{pink noise} from
white noise, which requires a filter rolling off $-3$ dB per octave.
Pink noise is also called ``$1/f$ noise'', referring to the roll-off
of the \emph{power spectral density} of the noise, which requires a filter
for white-noise having a magnitude response proportional to
$1/\sqrt{f}$.  Many natural processes have been found to be well
modeled by $1/f$ noise, such as amplitude fluctuations in classical
music, sun spots, the distribution of galaxies, transistor flicker
noise, flood levels of the river Nile, and more
\cite{VossAndClarke78}.\ftu{http://123.physics.ucdavis.edu/week_3_files/voss-clarke.pdf}

The ideal filter for synthesizing $1/f$ noise from white noise has transfer function
\[
H_{-\frac{1}{2}}(s) = \frac{1}{\sqrt{s}} \;\goestoas{s=j\omega}\; \frac{1}{\sqrt{j}\,\sqrt{\omega}} \eqsp e^{-j\frac{\pi}{4}}\, \omega^{-\frac{1}{2}},
\]
corresponding to $\alpha=-1/2$ in \eref{Halpha}.  Since the filter
phase is arbitrary when filtering white noise, the filter-design
problem can be formulated to match only the power frequency response
$|H_{-1/2}(j\omega)|^2 = 1/\omega$ (hence the name ``$1/f$ filters''),
thereby obtaining a distribution of poles and zeros yielding a
frequency response proportional to $1/\sqrt{\omega}$ for frequencies
$\omega=2\pi f$ in some finite range $f\in[\fmin,\fmax]$.  For audio,
we ideally choose $\fmin\approx 20$ Hz and $\fmax\approx 20$ kHz.
Such designs can be found on the
Web\ftu{https://ccrma.stanford.edu/~jos/sasp/Example_Synthesis_1_F_Noise.html}
and in the \Faustsp
distribution.\ftu{faust.grame.fr} There are also
interesting ``Voss-McCartney algorithms'' which are essentially sums
of white-noise processes that are sampled-and-held at various
rates.\ftu{http://www.firstpr.com.au/dsp/pink-noise/}

\subsection{Summary of Results}

In this paper we derive closed-form expressions for the poles and
zeros of spectral roll-off filters having any desired slope to a
controllable degree of accuracy. The accuracy desired and the
bandwidth over which the approximation holds dictate the order of the
filter required, but the basic structure of the filter never varies.
The poles and zeros are all real, and they alternate, with
exponentially increasing spacing (uniformly spaced on a log scale).  A
simple initial derivation can be based on \emph{Bode Plot} analysis,
as described in the next section.

\subsection{Outline of the Remainder}

We first review Bode magnitude plots, and then design $f^{\alpha}$
filters accordingly.  We then evaluate the quality of the
approximation, and develop a practical design algorithm.
Discrete-time filter design using the bilinear transform with
frequency prewarping is discussed.  Finally, software implementations
are given in the matlab and \Faustsp languages.

\section{Bode Plots}

A \emph{Bode Plot} of a filter frequency response $H(j\omega)$
separately graphs the log-magnitude and phase versus
log-frequency.\ftu{https://en.wikipedia.org/wiki/Bode_plot}
We are only concerned here with log-magnitude plots, and will omit
consideration of the Bode phase plot, which happen to behave as
expected naturally.  The usual choice of log-magnitude units is
\emph{decibels} (dB) $20\log_{10}\left[|H(j\omega)|/R\right]$
(relative to an arbitrary reference, such as $R=1$), and the
log-frequency axis is typically either in \emph{octaves}
($\propto\log_2(\omega)$) or \emph{decades}
($\propto\log_{10}(\omega)$).  Thus, a single pole is said to give a
roll-off of $-6$ dB per octave or $-20$ dB per decade.  Octaves are
typical in audio signal processing while decades are typical in the
field of automatic control.


\begin{figure}[ht]
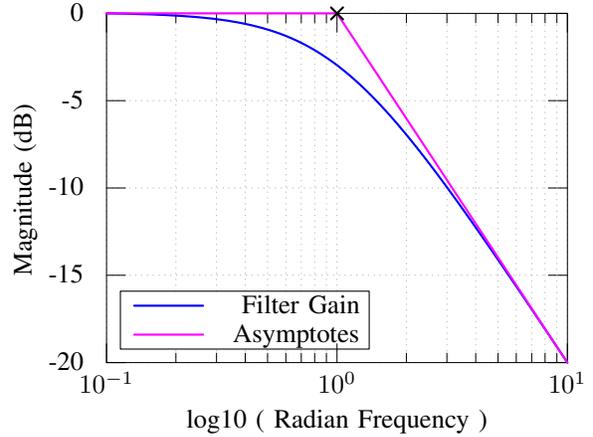

\input ./tikz/bode-one-pole.tikz
\caption{Bode plot and its asymptotes (``stick
  diagram'') for a single pole at $s=-1$, \ie, the transfer function
  $H(s)=1/(s+1)$.}
\label{fig:bode-one-pole}
\end{figure}

\Fref{bode-one-pole} illustrates the Bode plot and its associated
``stick diagram'' (comprised of asymptotic gains) for a single pole at
$s=-1$.  We see that the response is flat for low frequencies, drops
to $-3$ dB at the break frequency $\omega=1$, and approaches the $-20$
dB per decade asymptote, reaching the asymptote quite well by one
decade above the break frequency at $\omega=10$.

For a general filter transfer function having $N$ poles $p_n$ and $M$ zeros $z_m$
\begin{equation}
H(s) = g\frac{\prod_{m=1}^M (s-z_m)}{\prod_{n=1}^N (s-p_n)}
\label{e:H}
\end{equation}
the Bode plot can be expressed as
\[
\Bt(\wt) = g_1 \log_{b_1} H\left(jb_2^{\log_{b_2}(\omega)}\right) = \Gt(\wt) + j\Theta(\wt),
\]
where $g_1=20$, $b_1=10$, $\wt=\log_{b_2}(\omega)$, and $b_2$ is typically $2$ or $10$.
For mathematical simplicity, however, we'll choose instead $g_1=1$ and $b_1=b_2=e$,
giving
\[
\Bt(\wt) = \ln H\left(je^\wt\right) = \Gt(\wt) + j\Theta(\wt),
\]
where $\wt = \ln(\omega)$.  In this choice of units, $N$ integrators
$H(s)=1/s^N$ give a magnitude roll-off of $-N$ ``nepers per neper'',
while $M$ differentiators $H(s)=s^M$ give a slope of $+M$ in the Bode
magnitude plot
\begin{equation}
\Gt(\wt) = \realpart{\Bt(\wt)} = \realpart{\ln H\left(je^\wt\right)}.
\label{e:Gt}
\end{equation}

Our problem is to find poles and zeros of $H(s)$ to minimize some norm of the error
\[
\left\|W(\wt)\left[\Gt'(\wt) - \alpha \right]\right\|
\]
where $\Gt'(\wt)=d\,\Gt(\wt)/d\wt$ denotes the derivative of $\Gt(\wt)$ with respect to
$\wt$, $\alpha$ is the desired slope of the log-magnitude
frequency-response $\Gt'$ versus log frequency $\wt$, and $W(\wt)$
denotes a real, nonnegative weighting function.

As a specific example, for the Chebyshev norm and a uniform weighting
$W(\wt)\equiv 1$ between frequencies $\wt_1$ and $\wt_2$, we have
\[
\min_H\left\{\max_{\wt\in[\wt_1,\wt_2]} \left|\Gt'(\wt) - \alpha \right|\right\}.
\]
That is, we wish to minimize the worst-case deviation between the
desired slope $\alpha$ and the achieved slope $\Gt'(\wt)$ over a
specific (audio) band $[\wt_1,\wt_2]$.

\section{Stick Diagram Design}

When designing a filter with a prescribed magnitude response by the
Bode ``stick diagram'' method, we think in terms of poles and zeros
``breaking'' at certain frequencies.  For example, in the term
\[
H_n(j\omega) = \frac{-p_n}{j\omega-p_n}
\]
which is scaled to have unity gain at $\omega=0$, the pole is said to
``break'' at frequency $\omega=|p_n|$.  This is easily seen to be the
$-3$ dB point of the term, since
\[
H_n(jp_n) = \frac{-p_n}{j (\pm p_n)-p_n} = \frac{1}{1\pm j}
\]
which has magnitude $1/\sqrt{2} \approx -3$ dB.  Thus, the gain of the
term is approximately constant out to $\omega=|p_n|$, where it reaches
its $-3$ dB, or ``half power'' frequency, followed by its asymptotic
roll-off of $-6$ dB per octave.  A zero term $H_m(j\omega) =
(j\omega-z_m)/(-z_m) = 1-j\omega/z_m$ similarly starts out flat,
reaches magnitude-gain $+3$ dB at its break frequency $\omega=|z_m|$,
and asymptotically approaches $+6$ dB per octave for $\omega\gg
|z_m|$.

The Bode design procedure is then to start at dc ($\omega=0$) and map
out break-frequencies for poles and zeros so as to follow the desired
response as closely as desired.  Since this tool is commonly applied
in control-system design, there is also usually consideration for the
phase plot as well, which has similarly simple
behavior.\ftu{https://en.wikipedia.org/wiki/Phase_margin}

The basic Bode ``stick diagram'' consists only of straight line
segments, each having slope given by some integer number of nepers per
neper (or integer multiple of $\pm 6$ dB/octave, etc.), with the
knowledge that the actual response will be a smoothed version of the
stick diagram, traversing the $\pm 3$ dB points at line-segment
intersections corresponding to isolated breaking zeros and poles,
respectively.

\section{Approximating Arbitrary Slopes}

To approximate arbitrary slopes $\alpha$, we may alternate poles and
zeros so that the average slope of the stick diagram equals $\alpha$.

For example, to achieve a slope of $\alpha = -1/2$ (``half an
integrator''), we may start with a pole near $p_1=-2\pi f_1$, where
$f_1$ is our lowest frequency of interest (nominally $20$--$40$ Hz for
audio), which causes the slope to approach $-1$.  Then, half an octave
to the right, \eg, we can locate a zero to cancel the pole's roll-off,
pushing the slope from $-1$ back toward $0$.  Continuing in this way,
we may locate a pole at each octave point $f_12^k$, $k=0,1,2,\ldots\,$,
with zeros interlacing at $f_12^{k+1/2}$, in order to achieve an
\emph{average slope} of $\alpha = -1/2$.

\begin{figure}[ht]
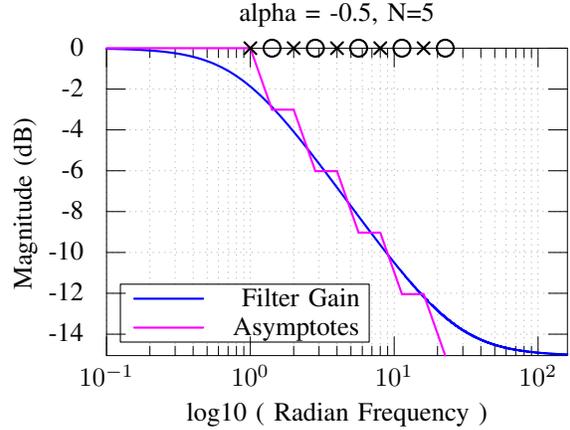

\input ./tikz/bode-pink-mp5-N5.tikz
\caption{Bode plot and its asymptotes (``stick
  diagram'') for $N=5$ poles and zeros, arranged to approximate a
  $1/f$ power response.}
\label{fig:bode-pink-mp5-N5}
\end{figure}

\Fref{bode-pink-mp5-N5} shows the Bode plot and the corresponding
stick diagram for $N=5$ poles located to give breakpoints distributed
along octaves starting with $\omega=1$.  That is, the poles are at
$s=p_n=-r^n$, for $r=2$ and $n=0,2,\ldots,N-1$.  To approximate a
$1/f$ power-response having slope $\alpha=-1/2$ nepers per neper, we
place $N$ zeros at $s=z_m=p_mr^{-\alpha}$, $m=0,2,\ldots,N-1$, \ie,
shifted half an octave toward higher frequency, interlacing the poles.
We see that the response is flat for frequencies below the first
break-frequency $\omega=1$ as before, but the gain drops by less the
$2$ dB at $\omega=1$ due primarily to the influence of the upcoming
first zero at $\omega=\sqrt{2}$. Between the pole-zero frequencies
$\omega=1$ and $\omega=r^{N-1-\alpha}$, the Bode plot smoothly
interpolates the stick diagram which alternates between slopes of $0$
and $-20$ dB per decade, as the poles and zeros break in alternation,
yielding an average roll-off of $-10$ dB per decade, as desired.  After
the last zero breaks, the response levels off to a final slope of $0$.
Alternatively, the last zero could be omitted to have a final $-20$ dB
per decade slope, etc.  Additionally, we plot the pole symbol `X' and
zero symbol `O' along the upper horizontal axis at their corresponding
break frequencies.  This plot suggest that we may be able to choose
$N$, $r$, and $p_1$ to achieve any desired accuracy over any finite
band.

More generally, for any desired slope $\alpha\in(-1,1)$, we place the
$k$th zero on the negative-real axis of the complex $s$ plane at $s =
z_k = p_kr^{-\alpha}$, $k=0,1,2,\ldots,N-1$, where $p_k=p_1r^k$
denotes the $k$th pole, exponentially distributed along the
negative-real axis with spacing ratio $r=p_{k+1}/p_k$, starting at
$p_1 = -2\pi f_1$.

Note that $\alpha=0$ cancels all of the poles with zeros, yielding a
constant magnitude frequency response, while $\alpha=-1$ cancels all
poles except the first $p_1$, leaving a slope of $-1$ nepers per neper
(an integrator) for $\omega\gg \omega_1$.  When $\alpha>0$, the
pole-zero sequence starts out from the origin of the $s$ plane with a
zero, thereby giving a net positive slope to the Bode magnitude plot.
In particular, at $\alpha=1$, all of the poles are canceled by zeros,
leaving behind a single zero at $s=p_1/r$, yielding a slope of $+1$
(differentiator) for $\omega\gg \omega_1/r$.  Between these extremes,
the poles and zeros interlace asymmetrically to approximate any
desired slope $\alpha$.

As examples of other slopes, \fref{bode-pink-p5-N5} shows a Bode plot
analogous to \fref{bode-pink-mp5-N5} for the case $\alpha=+1$ (``half
a differentiator''), and \fref{bode-pink-mp2-N5} shows $\alpha=-0.2$,
showing the resulting asymmetric pole-zero layout on a log-frequency
scale.

\begin{figure}[t]
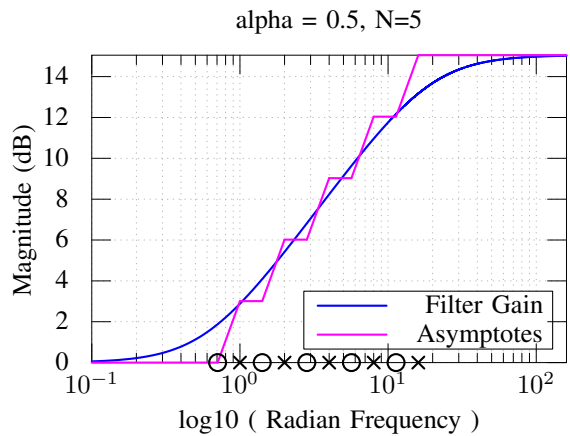

\input ./tikz/bode-pink-p5-N5.tikz
\caption{Bode plot and its asymptotes for $N=5$ poles
  and zeros, arranged to approximate ``half a differentiator''
  $|H_{1/2}(j\omega)|=\sqrt{\omega}$.}
\label{fig:bode-pink-p5-N5}
\end{figure}

\begin{figure}[t]
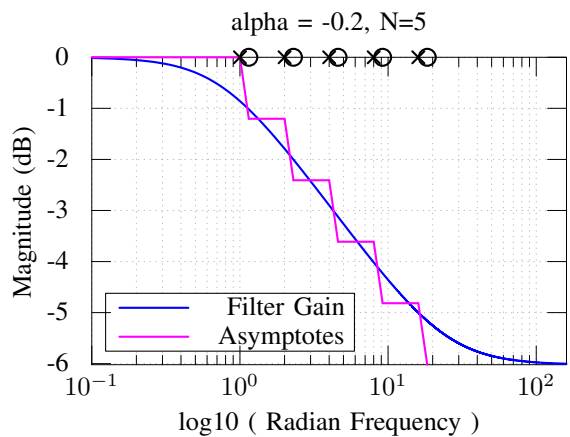

\input ./tikz/bode-pink-mp2-N5.tikz
\caption{Bode plot and its asymptotes for $N=5$ poles
  and zeros, arranged to yield a Bode plot with slope $\alpha=0.2$ over
  a chosen frequency range: $|H_{-0.2}(j\omega)|=\omega^{-0.2}$.}
\label{fig:bode-pink-mp2-N5}
\end{figure}

To reduce the maximum error, the interlacing pole-zero pattern can be
made more dense, \eg, by placing poles every half octave, or third
octave, etc.  As an example, \fref{bode-pink-mp5-N12} shows the
improvement obtained over \fref{bode-pink-mp5-N5} by increasing the
order from $N=5$ to $12$.

\begin{figure}[ht]
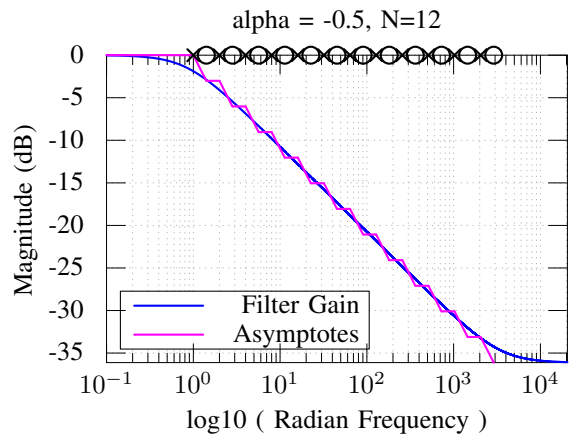

\input ./tikz/bode-pink-mp5-N12.tikz
\caption{Bode plot and its asymptotes (``stick
  diagram'') for $N=12$ poles and zeros, arranged to approximate a $1/f$ filter response.}
\label{fig:bode-pink-mp5-N12}
\end{figure}


It is not necessary for the slope $\alpha$ of the spectral roll-off to
be restricted between $-1$ and $1$ neper per neper.  Any number of
poles or zeros can be used in the low-frequency region to establish
any integer part for the slope, or some number of the regularly spaced
poles (zeros) can be skipped before the partial cancellation array
of zeros (poles) begins.  The subsequent interlacing poles and zeros
then only need to set the fractional value of the slope above $f_1$.

\begin{figure}[t]
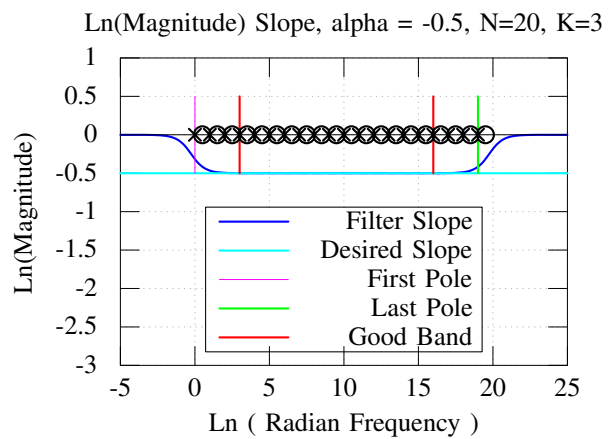

\input ./tikz/bode-slope-mp5-N20-K3.tikz
\caption{%
Slope-matching performance for 20 exponentially distributed pole-zero
pairs approximating a $1/f$ filter response ($\alpha=-1/2$), with the
first and last three poles placed outside the band of interest (``good
band'').}
\label{fig:bode-slope-mp5-N20-K3}
\end{figure}

\Fref{bode-slope-mp5-N20-K3} shows the quality of approximation to the
log-magnitude \emph{slope} for $\alpha=-1/2$ and $N=20$ pole-zero
sections.  The poles are arbitrarily located at $s=-e^n$, for
$n=0,1,2,\ldots,N-1$, yielding log-break-frequencies at
$\ln(\omega_n)=n$.

\Fref{bode-error-mp5-N20-K0} on page \pageref{fig:bode-error-mp5-N20-K0}
shows the log-magnitude slope \emph{error} for the example
of \fref{bode-slope-mp5-N20-K3}.  We see that the error is large
toward the edges of the pole-zero range.  This suggests that we may
define a new parameter $K$ which is the number of pole-zero sections
to \emph{skip} at the beginning and end of the pole-zero array so that
the error is much smaller over the band of
interest.  \Fref{bode-slope-mp5-N20-K3} marks the band defined by
$K=3$ using a second pair of vertical lines having the same color (see
legend).

\Fref{bode-error-mp5-N20-K3} shows the same slope error 
as \fref{bode-error-mp5-N20-K0} but for $K=3$.  Zoomed in like this,
the error plot strongly suggests that, for an infinite array of
pole-zero pairs, the optimal Chebyshev slope approximation is obtained
in the limit.

\clearpage

\begin{figure*}[ht]
\includegraphics[width=0.8\textwidth]{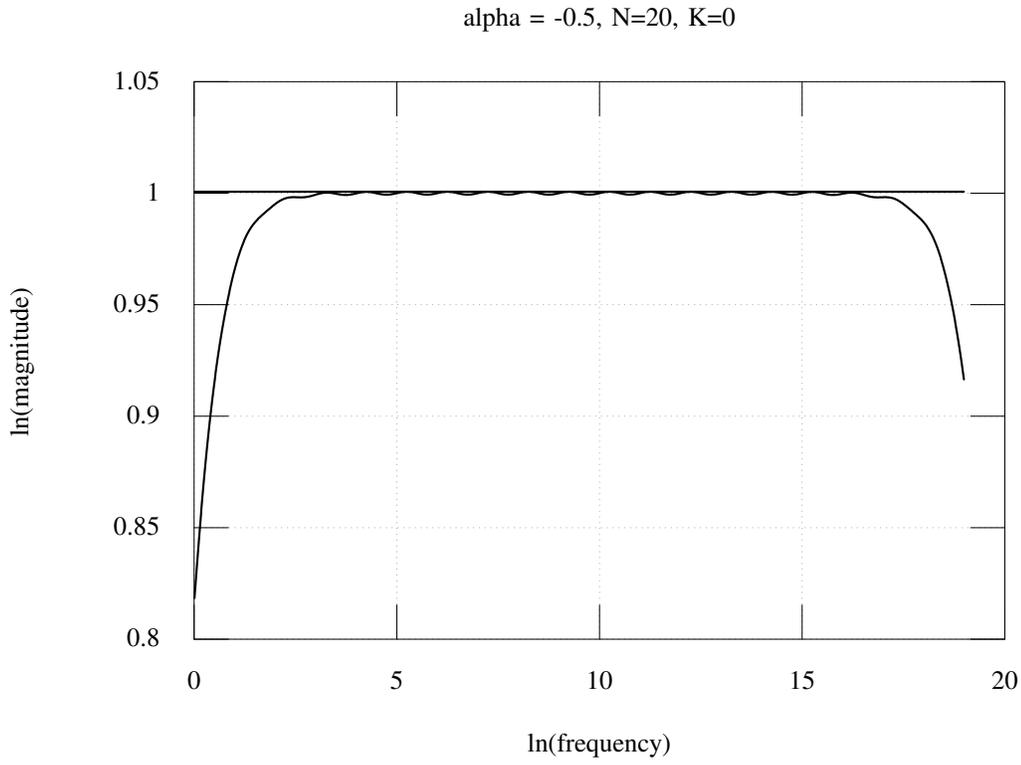}
\caption{Slope error when pole-zero pairs are used only across the band of interest.}
\label{fig:bode-error-mp5-N20-K0}
\end{figure*}

\begin{figure*}[ht]
\includegraphics[width=0.8\textwidth]{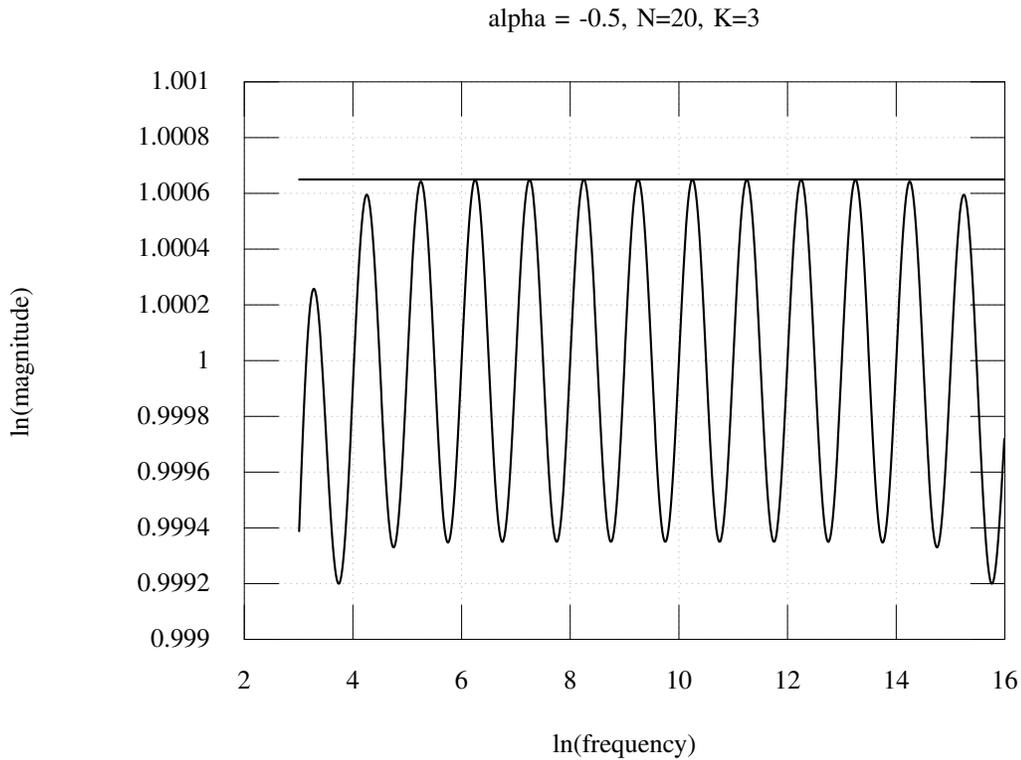}
\caption{Slope error when $K=3$ pole-zero pairs are placed both before and after the band of interest.}
\label{fig:bode-error-mp5-N20-K3}
\end{figure*}

\clearpage

\section{Mathematical Approximation}

As in \eref{Gt}, our Bode magnitude plot can be expressed as the real part of
\begin{equation}
\Bt(\wt) = \Ht\left(je^\wt\right) = \ln H\left(je^\wt\right)
\label{e:Gtt}
\end{equation}
and our problem is to find poles and zeros of $H(s)$ such that
\[
\Gt(\wt) \to \alpha\,\wt.
\]
for $\alpha\in(-1,1)$.  It is convenient to \emph{differentiate} with respect to $\wt$ and formulate the problem as
\[
\frac{d}{d\wt}\Gt(\wt) \to \alpha.
\]
Substituting the definition of $H(s)$ from \eref{H} into \eref{Gtt} yields
\begin{eqnarray*}
\Bt(\wt) &=& \Ht\left(je^\wt\right)\\[5pt]
	 &=& \ln(g) + \left.\left[\sum_m \ln(j\omega-z_m) - \sum_n (j\omega-p_n)\right]\right|_{\omega=e^\wt}\\[5pt]
         &=& \ln(g) + \sum_m \ln(je^\wt-e^{\zt_m}) - \sum_n (je^\wt - e^{\pt_n})
\end{eqnarray*}
and differentiating with respect to $\wt$ yields
\begin{eqnarray*}
\Bt^\prime(\wt) &=& \Ht^\prime\left(je^\wt\right)\\
&=& \sum_m \frac{je^{\wt}}{je^{\wt}-e^{\zt_m}}
  - \sum_n \frac{je^{\wt}}{je^{\wt}-e^{\pt_n}}\\
&=& \sum_m \frac{1}{1+je^{(\zt_m-\wt)}}
  - \sum_n \frac{1}{1+je^{(\pt_n-\wt)}}\\
&=& \sum_m \frac{1}{1+je^{\ln(z_m/\w)}}
  - \sum_n \frac{1}{1+je^{\ln(p_n/\w)}}\\
&=& \sum_m \frac{1}{1+j(z_m/\w)}
  - \sum_n \frac{1}{1+j(p_n/\w)}\\
&=& \zbox{\sum_m \frac{j\w}{j\w-z_m}
  - \sum_n \frac{j\w}{j\w-p_n}}\\
&=& \sum_m \frac{je^\wt}{je^\wt-e^{\zt_m}}
  - \sum_n \frac{je^\wt}{je^\wt-e^{\pt_n}}
\end{eqnarray*}
The analysis so far has carried along both log-magnitude and phase,
since it was no extra work to do so.  Taking now the real part to look
only at log-magnitude gives
\begin{eqnarray}
\Gt^\prime(\wt) &=& \realpart{\Ht^\prime\left(je^\wt\right)}\nonumber\\
&=& \realpart{\sum_m \frac{j\w}{j\w-z_m}
  - \sum_n \frac{j\w}{j\w-p_n}}\nonumber\\
&=& \realpart{\sum_m \frac{j\w(-j\w-z_m)}{|j\w-z_m|^2}
  - \sum_n \frac{j\w(-j\w-p_m)}{|j\w-p_n|^2}}\nonumber\\
&=& \realpart{\sum_m \frac{\w^2-jz_m\w}{\w^2+z_m^2}
            - \sum_n \frac{\w^2-jp_n\w)}{\w^2+p_n^2}}\nonumber\\
&=&
  \sum_m \frac{\omega^2}{\omega^2 + z_m^2}
- \sum_n \frac{\omega^2}{\omega^2 + p_n^2}\elabel{squares}\\
&=&\sum_m \frac{e^{2\wt}}{e^{2\wt}+e^{2\zt_m}}
- \sum_n \frac{e^{2\wt}}{e^{2\wt}+e^{2\pt_n}}\nonumber\\
&=&\sum_m \frac{1}{1+e^{2(\zt_m-\wt)}}
 - \sum_n \frac{1}{1+e^{2(\pt_n-\wt)}}.
\end{eqnarray}
where, thanks to the squaring of all terms in \eref{squares}, we may
consider only the absolute values of the frequency $\omega$, zeros
$z_m$, and poles $p_n$:
\begin{eqnarray*}
\wt &=& \ln(|\omega|)\\
\zt_m &=& \ln(|z_m|)\\
\pt_n &=& \ln(|p_n|)
\end{eqnarray*}

Thus, the normalized basic building block of the Bode log-magnitude plot is given by
\[
\Gt_{z_m}^\prime(\ln \w) = \frac{\w^2}{\w^2+z_m^2}
\]
for a zero $z_m$, and
\[
\Gt_{p_n}^\prime(\w) = -\frac{\w^2}{\w^2+p_n^2}
\]
for a pole $p_n$.

Let $b$ denote either a pole or a zero (the ``break frequency'' in
rad/s due to the pole or zero):
\[
\Gt_b^\prime(\ln \w) = \pm\frac{\w^2}{\w^2+b^2}
\]
For $|\w|\ll b$, the slope is zero,
while for $|\w|\gg b$, it approaches a slope of $+1$ for a zero and
$-1$ for a pole.  At $|\w|=b$, the slope is $\pm 1/2$.

In log-frequency units, our basic slope building-block can be written
\[
\Gt_b^\prime(\wt) \eqsp \pm\frac{e^{2\wt}}{e^{2\wt}+e^{2\bt}} \eqsp \pm\frac{1}{1+e^{-2(\wt-\bt)}}.
\]
We may \emph{normalize} this building block to the case $b=0$
\[
\Gt_0^\prime(\wt) \eqsp \pm\frac{1}{1+e^{-2\wt}}
\]
so that $\Gt_b^\prime(\wt) = \Gt_0^\prime(\wt-\bt)$. That is, the
slope of our desired log-magnitude frequency response is given by
sums and differences of arbitrary shifts of $\Gt_0^\prime$.

The ``stick diagram'' approximation of our slope basis function is
given by
\[
\Gt_0^\prime(\wt) \;\approx\; \pm u(\wt),
\]
where $u(t)$ denotes the Heaviside unit step function, which steps
from 0 to 1 as $t$ goes from negative to positive.  Thus, each zero
$\zt_n$ of the transfer function approximately contributes a
positive-going step $u(\wt-\zt_n)$ to the sum of terms making up the
overall slope, while each pole $\pt_m$ contributes $-u(\wt-\pt_m)$

\subsection{Pulse Train View of the Approximation}

It is well known that a rectangular pulse $\pulse_\tau(t)$ of width $\tau$
can be synthesized from two unit steps $u(t)$ as
\[
\pulse_\tau(t) = u(t) - u(t-\tau).
\]
We can furthermore create a periodic pulse train with period $t_0$ by
means of periodic replication of $\pulse_\tau$:
\[
\shah(t;\tau,t_o) = \sum_{n=0}^{\infty}\pulse_\tau(t-nt_0)
\]
The mean of this periodic rectangular-pulse train is $\alpha =
\tau/t_0$, adjustable between 0 and 1.  Note that
$\alpha$ is also the \emph{duty cycle} of the pulse train (ratio of
pulse width $\tau$ to period $t_0$).

\section{Pole-Zero Placement Algorithms}

Applying this ``Bode thinking'' to the fractional slope problem, we may
\begin{enumerate}
\item choose an exponential pole spacing $\Delta_p = |\pt_{k+1}| - |\pt_k|$
  that will eventually determine our filter order, and
\item set the pole-to-zero spacing
\[
\Delta_z = |\zt_k| - |\pt_k|
\]
so as to achieve the desired duty cycle
\[
\alpha = -\frac{\Delta_z}{\Delta_p},
\]
where the minus sign appears because we wish $z_k<p_k<0$ which gives a
negative slope, thereby effectively choosing
$-\shah(\wt;\Delta_z,\Delta_p)$ as our stick-diagram pulse train.
\end{enumerate}
We thus choose $\Delta_p$ arbitrarily according to how many poles we
can afford (or to yield a sufficiently small error), and set
\begin{eqnarray*}
  \Delta_z &=& -\alpha \Delta_p \eqsp -\alpha \cdot (|\pt_{k+1}| - |\pt_k|)\\[5pt]
  \Rightarrow \qquad \zt_k & = & \pt_k - \alpha \Delta_p.
\end{eqnarray*}

For example, choosing octave spacing for the poles corresponds to $\Delta_p=\ln(2)$.

Defining
\[
r = \frac{p_{k+1}}{p_k},
\]
we may write
\[
p_k = p_0 r^k, \quad k\in \ints
\]
where $p_0<0$ may be arbitrarily chosen on the negative-real axis, and
\[
z_k = p_k r^{-\alpha}.
\]
We may now assemble the complete transfer function as
\[
H_\alpha(s; r, p_0) = \prod_{k=-\infty}^\infty\frac{s-p_0r^{k-\alpha}}{s-p_0r^k}.
\]
We conjecture that $\lim_{r\to1}H_\alpha(j\omega; r,p_0) = e^{j\alpha\frac{\pi}{2}} \omega^\alpha$, \ie,
\beqn
\zbox{\lim_{r\to1}\prod_{k=-\infty}^\infty\frac{j\omega-p_0r^{k-\alpha}}{j\omega-p_0r^k} \eqsp e^{j\alpha\frac{\pi}{2}} \omega^\alpha}
\elabel{conj}
\eeqn
for any real $p_0<0$, \ie, that the approximation becomes exact in the
limit as the pole-zero density goes to infinity in this configuration.


\subsection{Chebyshev Optimality Argument}

The log-magnitude approximation error in \eref{conj} is a periodic
function of $\ln(\omega)$ and is therefore an ``equal-ripple''
oscillating error, as required for Chebyshev optimality.  Furthermore,
each pole-zero pair corresponds to one cycle of this oscillation, with
each pole creating a specific negative error-peak, and each zero a
specific positive error-peak. Thus, the correspondence between
error-peaks and poles-and-zeros is bijective.  Considering the degrees
of freedom in the magnitude-response of a collection of poles and
zeros, there can be no other error peaks.  Therefore, the Chebyshev
optimal configuration must be as found, provided that the basic
Chebyshev approximation theorems hold in this extended setting.

\subsection{Truncated Pole-Zero Pairs}

In practice, we must choose a specific frequency interval
$[\fmin,\fmax]$ outside of which we do not require a $f^\alpha$
response.  We then have a trade-off between the number of poles $N$
required and the approximation error between $\fmin$ and $\fmax$.  We
may choose an error tolerance and determine the number of poles
required, or simply minimize the error for a given number of poles.

\subsection{Specifying Filter Order and Band of Interest}

It is straightforward to solve for the pole-placement ratio $r$ and
first pole-frequency $f_1$ given the desired filter order $N$ and the
band of interest $[\fmin,\fmax]$:
\begin{enumerate}
\item Specify the desired number of poles $N$ and approximation frequency-range $[\fmin,\fmax]$ Hz.
\item Determine the first pole frequency $f_1$ and pole ratio $r=f_{k+1}/f_k$ by solving
\[
\twobytwo{1}{K}{1}{N-K-1} \twobyone{\ft_1}{\rt} \eqsp \twobyone{\ftmin}{\ftmax}
\]
for $\ft_1$ and $\rt$, where $\tilde{x}$ denotes $\ln(x)$.  The
integer $K$ is the number of poles occurring before and after the
frequency range $[\fmin,\fmax]$.  Numerical experiments indicate that
$K=3$ is a cost-effective choice, with higher values yielding somewhat
smaller error in the desired frequency range $[\fmin,\fmax]$.
\item Then the $n$th pole is given by
\[
p_n \eqsp -2\pi f_1 r^{n-1}
\]
for $n=1,2,\ldots, N$.  The $m$th zero is
\[
z_m \eqsp p_m r^{-\alpha}
\]
where $\alpha\in[-1,1]$ is the desired slope of the Bode magnitude
plot (in $\ln$/$\ln$ units).  
\end{enumerate}
As mentioned above, any integer part of the slope can be obtained by
preceding the pole-zero array with the desired number of poles or
zeros.  While we could interchange the roles of poles and zeros to
change the sign of the slope, it is better in practice to leave the
poles fixed, and only vary the zeros to modulate the slope.  Zeros can
be dynamically modulated more robustly than poles, so real-time
modulation of the spectral slope is best carried out by sliding the
array of zeros uniformly to the left and right along the negative real
log axis in the $s$ plane.  Note that as the array of zeros crosses
the array of poles, an integer part is incremented or decremented in
the slope $\alpha$.  Thus, there is no practical restriction to
$\alpha\in(-1,1)$.


\section{Digitization}

The filters considered up to now are all for continuous-time
processing.  We look now at the effects of \emph{digitization} on the
accuracy of filter slope.


The \emph{bilinear transform}\ftu{https://ccrma.stanford.edu/~jos/pasp/Bilinear_Transformation.html}
digitizes a filter by means of the substitution
\begin{equation}
s \eqsp c\frac{1-\zi}{1+\zi} \;\;\Leftrightarrow\;\; z \eqsp \frac{1+s/c}{1-s/c}
\label{eq:Ebilin}
\end{equation}
where $c$ is some positive real constant. That is, given a continuous-time
transfer function $H_a(s)$, we apply the bilinear transform by defining
\[
H_d(z) = H_a\left(c\frac{1-\zi}{1+\zi}\right)
\]
where the ``$d$'' subscript denotes ``digital,'' and ``$a$'' denotes
``analog.''

Denoting the continuous-time radian-frequency axis by $\wa$ and the
the discrete-time radian-frequency axis by $\omega_d$, we find the frequency-warping
of the bilinear transform to be
\[
j\,\wa \eqsp c \,\frac{1-e^{-j\omega_d T}}{1+e^{-j\omega_d T}}
       \eqsp j \, c \, \tan\left(\frac{\omega_d T}{2}\right),
\]
where $T$ denotes the discrete-time sampling interval in seconds.
Thus, we may interpret $c$ as a \emph{frequency-scaling} constant.  At
low frequencies, $\tan(x)\approx x$, so that $\wa\approx c \omega_d T
/ 2$ at low frequencies, leading to the typical choice of $c = 2/T =
2f_s$, where $f_s$ denotes the sampling rate in Hz.  However, $c$ can be
chosen to map exactly any particular interior frequency
$\omega_a\in(0,f_s/2)$.

In our problem, we have complete control over the pole frequencies.
We may therefore \emph{prewarp} the pole locations so that they map to
an exact geometric progression.  We will therefore choose $c$ in the
bilinear transform to exactly map the break-frequency $f_1$ of our
first pole (or zero if starting out with a zero):
\[
c \eqsp \frac{2\pi f_1}{\tan(\pi f_1 T)}
\]
Next, we alter our $s$-plane pole frequencies $f_k$ for $k=2,3,\ldots$ to compensate
for the frequency-warping of the bilinear transform:
\[
\fh_k \eqsp c\,\frac{\tan\left(\pi f_k T\right)}{2\pi}
      \eqsp f_1\, \frac{\tan\left(\pi f_k T\right)}{\tan\left(\pi f_1 T\right)}
\]
where $\fh$ denotes the \emph{prewarped} version of $f$.  Since the
digital bandwidth is limited to half the sampling rate $f_s=1/T$, we
limit the number of digital-filter poles to $N$ for which $f_{N+1}\le
f_s/2$ while $f_{N+2}>f_s/2$. This choice leaves a full
pole-separation interval for the last zero to traverse as $\alpha$
traverses $[-1,1]$.

One characteristic of the bilinear transform is that it maps any poles
at infinity in the $s$ plane to $z=-1$ in the $z$ plane.  This means
we should either (1) choose the number of poles and zeros to be equal,
so that outside the band of interest the response levels off to a
constant magnitude, or (2) have more poles than zeros in the $s$-plane
filter so that the resulting zeros at infinity will map harmlessly to
$z=-1$ in the digital filter.  Poles at infinity, such as in the ideal
differentiator $H(s)=s$, give an unstable digital filter under the
normal bilinear transform.


\section{\Faustsp Implementation}

The function \texttt{spectral\_tilt} has been contributed to
\texttt{filter.lib} in the \Faust\ftu{http://faust.grame.fr}
distribution\footnote{Commit to master branch at
\url{ssh://git.code.sf.net/p/faudiostream/code/} on May 28, 2016}
having the following API:
\begin{lstlisting}
// USAGE:
//   _ : spectral_tilt(N,f0,bw,alpha) : _
// where
//     N = desired integer filter order 
//         (fixed at compile time)
//    f0 = lower frequency limit for 
//         desired roll-off band
//    bw = bandwidth of desired roll-off
// alpha = slope of roll-off desired 
//         in nepers per neper 
//         (ln mag / ln radian freq)
\end{lstlisting}

Also, the file \texttt{spectra\_tilt.dsp} was added to
the \Faustsp \texttt{examples} directory that
invokes \texttt{spectra\_tilt\_demo(N)} which is also defined
in \texttt{filter.lib}.

\section{Summary}

We have derived closed-form approximate fractional
integrator/differentiator filters $|H(j\omega)|=\omega^\alpha$ as
exponentially distributed real pole-zero pairs.  The approximation
error can be made arbitrarily small by reducing the spacing of the
pole-zero pairs and by extending them across a larger log-frequency
band than what is being used.  The poles are uniformly spaced along
the negative-real log-axis of the $s$ plane.  The zeros interleave the
poles and are spaced identically, but the relative spacing between the
array of poles and the array of zeros is $\alpha\in(-1,1)$.  Arbitrary
spectral slopes may be obtained dynamically by sliding the array of
zeros relative to the array of poles, without altering their internal
spacing.  The nature of the log-magnitude slope approximation
approaches Chebyshev optimality in the interior of the pole-zero
array, approaching conjectured Chebyshev optimality over all
frequencies in the limit as the order approaches infinity.  Software
implementations were provided in matlab and \Faust.

\section{Future Work}

We conjecture that in the case of an infinite pole-zero array, the
optimal Chebyshev $f^\alpha$ filter is obtained.  Therefore, it should
be possible to get closer to the Chebyshev approximation for finite
pole-zero intervals by addressing the ``edge effects'' due to array
truncation.  It is of practical interest to find compensation
strategies for these edge effects.  Otherwise, iterative methods can
be used as usual to convert the truncated pole-zero array toward the
optimal Chebyshev approximation.


\newcommand{\Appl}{Applications of}
\newcommand{\Trans}{Transactions on}
\newcommand{\Int}{International}
\newcommand{\InstOf}{Inst.\ of}
\newcommand{\Journal}{Journal of the}
\newcommand{\JournalOfThe}{Journal of the}
\newcommand{\JournalP}{Journal}
\newcommand{\Lab}{Laboratory}
\newcommand{\Eng}{Engineering}
\newcommand{\Elec}{Electrical}
\newcommand{\Soc}{Society}
\newcommand{\Proc}{Proceedings of the}
\newcommand{\Conf}{Conference}
\newcommand{\ConfOn}{Conference on}
\newcommand{\Tech}{Technical}
\newcommand{\Symp}{Symposium}
\newcommand{\Dept}{Department}
\newcommand{\Ama}{America}
\newcommand{\SocOf}{Society of}
\newcommand{\Acoust}{Acoustical}

{\raggedright
\bibliographystyle{IEEEtran}
\bibliography{stfh}
}


\clearpage


\begin{onecolumn}
\section*{Appendix: Matlab Software for Bode-Plot Figures}

\noindent
The following matlab was used to generate Figures
\ref{fig:bode-pink-mp5-N5} through \ref{fig:bode-error-mp5-N20-K3}:

{\small
\begin{lstlisting}[language=Matlab, firstline=5]
if !exist('dopause'), dopause=0; end

linewidth = 2;  % need 4 or more for EPS, 2 fine for TIKZ
bpline = '-b';  % Bode magnitude plot
asline = '--m'; % Bode magnitude plot asymptotes
% (dashed/dotted lines not available in tikz from (MacPorts) Octave right now,
%  so carefully choose colors to be distinguishable as gray levels)

% Bode magnitude plot for N-pole, N-zero filter approximating f^alpha

%N = 5; % Number of poles
N = 12; % Number of poles
alphas = [-1/2, 1/2, -0.2]; nalphas = length(alphas);
r = 2; % set closer to 1 for increased accuracy
for ia = 1:nalphas
  alpha = alphas(ia);
  mz = r .^ (-alpha+[0:N-1]); % minus the zeros
  mp = r .^ [0:N-1]; % minus the poles
  wa = [0.1*min(mp):0.01:10*max(mp)]; % radian frequency axis
  Ba = poly(mz) % filter transfer-function numerator B(s)
  Aa = poly(mp) % filter transfer-function denominator A(s)
  gainScale = Aa(end)/Ba(end); % Scale for unity dc gain (s plane)
  Ba = Ba * gainScale;
  Ha = freqs(Ba,Aa,wa);
  pdata = db(Ha); % filter magnitude frequency response in dB
  semilogx(wa,pdata,bpline,'linewidth',linewidth); grid; % Bode plot
  hold on % now draw the asymptotes:
  curDB = 0;
  pw = wa(1);
  slopeNeg = alpha<0;
  if slopeNeg, slope=-6.02; else slope=6.02; end
  for i=1:N
    if slopeNeg, bw=mp(i); else bw=mz(i); end  % Head to next break frequency
    semilogx([pw bw],[curDB curDB],asline,'linewidth',linewidth);
    pw = bw;
    % contine to next zero|pole at broken slope:
    if slopeNeg, bw = mz(i); else bw = mp(i); end
    newDB = curDB + slope*log2(bw/pw); % pole i breaks
    semilogx([pw bw],[curDB newDB],asline,'linewidth',linewidth);
    curDB = newDB;
    pw = bw;
  end
  semilogx([pw wa(end)],[curDB curDB],asline,'linewidth',linewidth);

  % Plot poles and zeros swung around to jw axis:
  semilogx(mz,zeros(1,length(mz)),'ok','linewidth',linewidth);
  semilogx(mp,zeros(1,length(mp)),'xk','linewidth',linewidth);
  axis tight;
  xlabel('log10 ( Radian Frequency )');
  ylabel('Magnitude (dB)');
  % title('Bode Plot and Stick Diagram');
  title(sprintf('alpha = %0.1f, N=%d',alpha,N));
  loc = 'southwest';
  if slope>0, loc = 'southeast'; end
  legend('Filter Gain','Asymptotes','location',loc);
  hold off
  if dopause, disp('PAUSING'); pause; end
  fname = bode_filename('bode-pink',alpha,N);
  print(['./tikz/',fname,'.tikz'],'-dtikz'); % for inclusion in LaTeX
end % for ia = 1:nalphas
\end{lstlisting}
}
\end{onecolumn}

\end{document}